\documentclass[twocolumn,aps,prl,preprintnumbers,bibnotes10pt,superscriptaddress]{revtex4}

\usepackage{amsmath}
\usepackage{graphicx}
\usepackage{ulem}
\usepackage{dcolumn}
\usepackage{epsfig}
\usepackage{bm}
\usepackage{array}
\usepackage[english]{babel}
\usepackage[pdfstartview=FitH, pdfstartpage=1, pdfpagemode=UseNone]{hyperref}
\hypersetup{
colorlinks=true,
citecolor=black,
linkcolor=black,
urlcolor=black
 , bookmarks=true,
 pdfmenubar=true
}

\usepackage{graphicx}
\usepackage{amsmath}
\usepackage{bm}           
\usepackage{ulem}

\newcommand{\ket}[1]{\ensuremath{|#1\rangle}}

\newcommand{\mean}[1]{\ensuremath{\big\langle #1 \big\rangle}}

\newcommand{\tr}{{\rm Tr}}

\newcommand{\vect}[1]{\bm{#1}}

\newcommand{\be}{\begin{equation}}
\newcommand{\ee}{\end{equation}}

\newcommand{\beq}{\begin{eqnarray}}
\newcommand{\eeq}{\end{eqnarray}}

\newcommand{\rhoT}{\hat{\rho}_T}
\newcommand{\chiT}{\chi_Q}
\newcommand{\chicl}{\chi_{\rm cl}}
\newcommand{\chimom}{\chi_{\rm mom}}
\newcommand{\Chimom}{$\chi_{\rm mom}$}
\newcommand{\Chicl}{$\chi_{\rm cl}$}

\begin{document}

\title{Adiabatic Sensing Enhanced by Quantum Criticality}
\author{Luca Pezz\`e}
\affiliation{QSTAR, Largo Enrico Fermi 6, 50125 Firenze, Italy} 
\affiliation{Istituto Nazionale di Ottica, INO-CNR, Via Carrara 1, 50019 Sesto Fiorentino, Italy}
\affiliation{LENS, Via Carrara 1, 50019 Sesto Fiorentino, Italy}
\author{Andreas Trenkwalder}
\affiliation{Istituto Nazionale di Ottica, INO-CNR, Via Carrara 1, 50019 Sesto Fiorentino, Italy}
\affiliation{LENS, Via Carrara 1, 50019 Sesto Fiorentino, Italy}
\affiliation{Dipartimento di Fisica e Astronomia, Universit\`a di Firenze, Via Carrara 1, 50019 Sesto Fiorentino, Italy}
\author{Marco Fattori}
\affiliation{LENS, Via Carrara 1, 50019 Sesto Fiorentino, Italy}
\affiliation{Dipartimento di Fisica e Astronomia, Universit\`a di Firenze, Via Carrara 1, 50019 Sesto Fiorentino, Italy}

\begin{abstract}
We propose an optimal method exploiting second order quantum phase transitions to perform high precision measurements of the control parameter at criticality. 
Our approach accesses the high fidelity susceptibility via the measurement of first- and second-moments of the order parameter and overcomes 
the difficulties of existing methods based on the overlap between nearby quantum states, which is hardly detectable in many-body systems.
We experimentally demonstrate the feasibility of the method with a Bose-Einstein condensate undergoing a symmetry-breaking quantum phase transition as a function of the attractive inter-particle interaction strength. 
Our moment-based fidelity susceptibility shows a clear peak that, at the same time, detects the quantum critical point at finite temperature without any model-dependent fit to the data and
certifies high sensitivity in parameter-estimation.
\end{abstract}

\maketitle
\date{\today}

Enhancing the sensitivity of measurements is a requirement to open new paths in science.
Gravitational wave detection, biological imaging, the search for variations of fundamental constants and measurement of time and gravity are 
forefront fields that require a constant search for new ideas and technologies aimed at decreasing the measurement uncertainties.
Sensors exploiting quantum technologies promise such 
a critical improvement in sensitivity~\cite{PezzeRMP2018, DeganRMP2017, SchnabelNATCOM2010, LudlowRMP2015, TaylorPR2016}.

Quantum sensors encode a physical quantity of interest into a parametric 
change of the system quantum state~\cite{DeganRMP2017, PezzeRMP2018}.
For instance, in an atomic or optical Mach-Zehnder interferometer, 
the input wavefunction acquires a relative phase shift proportional to an external field to be measured.
The more the state is susceptible to a parametric change, 
the easier is to distinguish it from neighbor states, and
the higher is the sensitivity.
The close relation between sensitivity and state susceptibility suggests~\cite{ZanardiPRA2008} to exploit the critical behavior of a many-body 
system close to a quantum phase transition (QPT)~\cite{SachdevBOOK,DuttaBOOK} to realize a new class of devices, 
which we call adiabatic sensors.
Approaching a QPT, the ground state of a many-body Hamiltonian $\hat{H}(\lambda)$ becomes
very susceptible to small changes of the control parameter 
$\lambda$~\cite{ZanardiPRE2006, YouPRE2007, ZanardiPRL2007, ZanardiPRA2008} 
(see \cite{BraunRMP2018, GuIJMPB2010} for reviews),
which is the physical quantity of interest.
At criticality, the sensitivity to the estimation of $\lambda$ can scale as
$(\Delta \lambda)^2 \sim 1/N^{2/(d\nu)}$ with the number of particles $N$~\cite{CamposVenutiPRL2007,SchwandtPRL2009,PolkovnikovBOOK}, 
where $d$ is the spatial dimension and $\nu$ is the critical exponent of the correlation length ($\xi \sim |\lambda - \lambda_c|^{-\nu}$).
On the contrary, ground states belonging to the same quantum phase can be poorly distinguished and $\Delta \lambda$ can reach, 
at best, a scaling $(\Delta \lambda)^2 \sim 1/N$~\cite{CamposVenutiPRL2007,SchwandtPRL2009,PolkovnikovBOOK}.  
It has been shown that several many-body models are characterized by 
$2/(d\nu) > 1$~\cite{GuIJMPB2010, BraunRMP2018, SchwandtPRL2009, PolkovnikovBOOK, ZanardiPRL2007, CamposVenutiPRL2007, RamsPRX2018, WangPRX2015, InvernizziPRA2008, RamsPRL2011, KwokPRE2008, SalvatoriPRA2014, BuonsantePRA2012, CozziniPRB2007, BuonsantePRL2007, MehboudiPRA2016} 
and can therefore be used as adiabatic quantum sensors with 
sensitivity $(\Delta \lambda)^2$ scaling faster than $1/N$ at criticality.  
However, these theoretical predictions have remained elusive so far because based on 
the analysis of the overlap between quantum states 
(and the related notion of quantum fidelity susceptibility~\cite{ZanardiPRE2006, GuIJMPB2010, BraunRMP2018, YouPRE2007, ZanardiPRL2007, CamposVenutiPRL2007, ZanardiPRA2008})
that is experimentally unfeasible in large systems~\cite{GuEPL2014, ZhangPRL2008, ZhangPRA2009}.

In this manuscript we overcome these limitations and demonstrate, for the first time, an adiabatic quantum sensor with 
sensitivity enhanced by quantum criticality.
We show that the optimal sensitivity predicted by the the quantum fidelity susceptibility
can be achieved by an experimentally-feasible parameter estimation method.
Differently from the existing proposals, this method does not require full access to the equilibrium state of the 
system but only the knowledge of first- and second-moments of the order parameter of the QPT.
We directly demonstrate the feasibility of the approach by analyzing experimental data obtained with a Bose-Einstein condensate trapped in a double-well potential
and undergoing a spontaneous symmetry breaking QPT as a function of the controllable inter-particle interaction~\cite{TrenkwalderNATPHYS2016, SpagnolliPRL2017}. 
The quantum critical point is recognized, even at finite temperature and
in presence of experimental noise, by a clear peak in the estimation sensitivity,
without any model-dependent assumption or theoretical fit to the data.
This provides a proof-of-principle direct demonstration that quantum criticality 
is a resource to enhance the estimation of the control parameter of a many-body Hamiltonian~\cite{ZanardiPRA2008, BraunRMP2018, footnote1}.

The experimental estimation of a parameter $\lambda$ is usually based on the detection of a 
$\lambda$-dependent mean value $\langle \hat{O} \rangle = \tr[\hat{\rho}(\lambda) \hat{O}]$ 
of some convenient measurement observable $\hat{O}$, where $\hat{\rho}(\lambda)$ is the quantum state of the sensor.
The sensitivity is obtained via error propagation $(\Delta \lambda)^2 = 1/\chimom(\lambda)$, where 
\be \label{chimom}
\chimom(\lambda) = \frac{1}{(\Delta \hat{O})^2} \bigg( \frac{d \mean{\hat{O}}}{d \lambda} \bigg)^2.
\ee
is a ``moment-based'' fidelity susceptibility and $(\Delta \hat{O})^2 = \tr[\hat{\rho}(\lambda) (\hat{O} - \mean{\hat{O}})^2]$ is the variance of $\hat{O}$.
Equation~(\ref{chimom}) fulfills the well known chain of inequalities~(see for instance Refs.~\cite{PezzeRMP2018,PezzeBOOK}) 
\be  \label{ineq}
\chi_{\rm mom}(\lambda) \leq \chicl(\lambda) \leq \chiT(\lambda),
\ee
valid for every quantum state and operator $\hat{O}$.
The left-side inequality expresses the fact that more
information on the parameter $\lambda$ can be captured by
high-order moments of the probability distribution $P(\mu \vert \lambda) = \langle \mu \vert \hat{\rho}(\lambda) \vert \mu \rangle$, 
where $\mu$ and $\ket{\mu}$ are eigenvalues and eigenstate of the operator $\hat{O}$ (namely, $\hat{O} \ket{\mu} = \mu \ket{\mu}$), respectively.
Here, 
\be \label{chicl}
\chicl(\lambda) =  -4  \frac{\partial^2 \mathcal{F}_{\rm cl}(\lambda,\epsilon)}{\partial \epsilon^2} \Big\vert_{\epsilon=0},
\ee
is a ``classical'' fidelity susceptibility (or Fisher information), where 
$\mathcal{F}_{\rm cl} (\lambda,\epsilon) = \sum_{\mu} \sqrt{P(\mu \vert \lambda) P(\mu \vert \lambda+\epsilon) }$
is the fidelity between probability distributions $P(\mu \vert \lambda)$ and $P(\mu \vert \lambda+\epsilon)$.
The corresponding sensitivity, $(\Delta \lambda_{\rm CR})^2 = 1/\chicl(\lambda)$, is the Cramer-Rao bound~\cite{CramerBOOK, HelstromBOOK}.
When optimizing $\chicl(\lambda)$ over all possible measurement observables $\hat{O}$, we obtain a fundamental bound of parameter estimation, 
called the quantum Cramer-Rao bound, $(\Delta \lambda_{\rm QCR})^2 = 1/\chiT(\lambda)$~\cite{HelstromBOOK, HelstromPLA1967, BraunsteinPRL1994}. The quantity
\be \label{chiL}
\chiT(\lambda) = \max_{\{ \hat{O} \}} \chicl(\lambda) =- 4 \frac{\partial^2 \mathcal{F}_Q(\lambda,\epsilon) }{\partial \epsilon^2} \Big\vert_{\epsilon=0}
\ee
is a ``quantum'' fidelity susceptibility~\cite{ZanardiPRE2006, YouPRE2007, ZanardiPRL2007, CamposVenutiPRL2007, GuIJMPB2010, BraunRMP2018, ZanardiPRA2008} (or quantum Fisher information). 
Here, $\mathcal{F}_Q(\lambda,\epsilon) = \tr[ ( \rho(\lambda)^{1/2} \rho(\lambda + \epsilon) \rho(\lambda)^{1/2} )^{1/2} ]$
is the Uhlmann fidelity between quantum states~\cite{UhlmannRMP1976, JozsaJMO1994}, which reduces to the 
overlap between ground states $\mathcal{F}_Q(\lambda,\epsilon) = \vert \langle \psi_0(\lambda + \epsilon) \vert \psi_0(\lambda) \rangle \vert$ 
at zero temperature.
$\chiT(\lambda)$ requires evaluating the Uhlmann fidelity between quantum states $\rho(\lambda)$ and $\rho(\lambda+\epsilon)$ that can be accessed 
via full state tomography that is experimentally unfeasible for large systems.
Alternatively, Eq.~(\ref{chiL}) shows that $\chiT(\lambda)$ coincides with a classical fidelity susceptibility for an optimal measurement observable $\hat{O}$ that, 
however, is often-impractical~\cite{nota5}. 

Here, we demonstrate that, for adiabatic sensing close to second-order QPTs, 
the optimal choice of measurement observable $\hat{O}$ in Eq.~(\ref{chimom}) is given by the order parameter of the QPT,
which can be easily accessible. 
Indeed, we find
\be \label{Eq}
\chi_{\rm mom}(\lambda) \sim \chicl(\lambda) \sim \chiT(\lambda)
\ee
valid for $\lambda$ around the critical point $\lambda_c$ and sufficiently small temperature $T$. 
Equation~(\ref{Eq}) provides a link between the 
Ginzburg-Landau approach to quantum critical phenomena -- where the symmetry breaking associated to the QPT is detected by a sudden variation of the order parameter 
and/or an increase of its fluctuations -- and the fidelity susceptibility approach -- which detects the QPT by a sudden change of the overlap between nearby quantum states.
To demonstrate Eq.~(\ref{Eq}) we exploit standard scaling relations and 
the sign ``$\sim$'' stands for equality up to a constant factor that cannot be determined with scaling arguments.
Let us consider a system composed of $N = L^d$ quantum spins, where $L$ is the linear system size.
A symmetry-breaking second order QPT is detected by a local order parameter $\hat{o}_i$
whose mean value scales at criticality as 
$\mean{\hat{o}_i} \sim (\lambda_c - \lambda)^\beta$ for $\lambda < \lambda_c$, where $\beta$ is (by definition) the order-parameter critical exponent.
The scaling behaviors of $\chi_{\rm mom}(\lambda)$, Eq.~(\ref{chimom}), is
obtained from the scaling of $\tfrac{d \langle \hat{O} \rangle}{d \lambda}$ and $\Delta^2 \hat{O}$, where $\hat{O} = \sum_{i=1}^N \hat{o}_i$ is a collective operator. 
We find $\tfrac{d \langle \hat{O} \rangle}{d\lambda} \sim \beta N  (\lambda_c - \lambda)^{\beta-1}$ and 
$(\Delta \hat{O})^2 \sim N (\lambda_c - \lambda)^{\nu z -\gamma}$, where 
$z$ is the dynamical critical exponent, $\nu$ is the correlation-length critical exponent and 
$\gamma$ is the susceptibility critical exponent~\cite{ContentinoBOOK}.
Combining these scaling relations, one obtains
\be \label{scalingchimom2}
\chi_{\rm mom}(\lambda)/N \sim (\lambda_c - \lambda)^{\nu d-2},
\ee
where we have used the quantum hyperscaling relation $2\beta + \gamma = \nu (d + z)$~\cite{ContentinoBOOK}.
For $\nu d < 2$, $\chi_{\rm mom}(\lambda)$ diverges, as a function of $\lambda$, at the critical point $\lambda_c$.
To find the scaling of $\chi_{\rm mom}(\lambda)$ with $N$ at criticality, 
we take into account that, for large but finite $N$, the critical point $\lambda_c^{(N)}$
approaches the asymptotic value $\lambda_c$ as
$(\lambda_c - \lambda_c^{(N)} ) \sim N^{-1/(d\nu)}$, 
for $\lambda_c^{(N)} < \lambda_c$~\cite{ContentinoBOOK}.
Combining this scaling law with Eq.~(\ref{scalingchimom2}), we obtain
\be \label{scalingchimom1}
\chi_{\rm mom}(\lambda_c^{(N)}) \sim \frac{1}{\Delta^2 \hat{O}\vert_{\lambda_c^{(N)}}} 
\bigg( \frac{d \mean{\hat{O}}}{d \lambda} \Big\vert_{ \lambda_c^{(N)}} \bigg)^2 \sim N^{2/(d\nu)},
\ee
a super-extensive scaling is found for $\nu d < 2$.
Equations (\ref{scalingchimom2}) and (\ref{scalingchimom1}) coincide with 
the known scaling behavior of $\chi_{\rm Q}$ as a function of $\lambda$ and $N$, respectively, see Refs.~\cite{CamposVenutiPRL2007,SchwandtPRL2009,PolkovnikovBOOK}.
Away from criticality, $\chi_{\rm mom}$ has, at best, an extensive scaling with $N$: this follows from the known
behavior of $\chi_{\rm Q}$~\cite{CamposVenutiPRL2007,SchwandtPRL2009,PolkovnikovBOOK} and the inequality $\chi_{\rm mom} \leq \chi_{\rm Q}$.
Following Eq.~(\ref{ineq}), the scaling behavior (\ref{scalingchimom1}) and (\ref{scalingchimom2}) hold also for $\chicl$. 
This demonstrates Eq.~(\ref{Eq}) and 
motivates the analysis of the fidelity susceptibility in a symmetry-breaking QPT using the experimentally-feasible $\chicl$ and $\chi_{\rm mom}$. 

\begin{figure*}[t!]
\begin{center}
\includegraphics[clip,width=\textwidth]{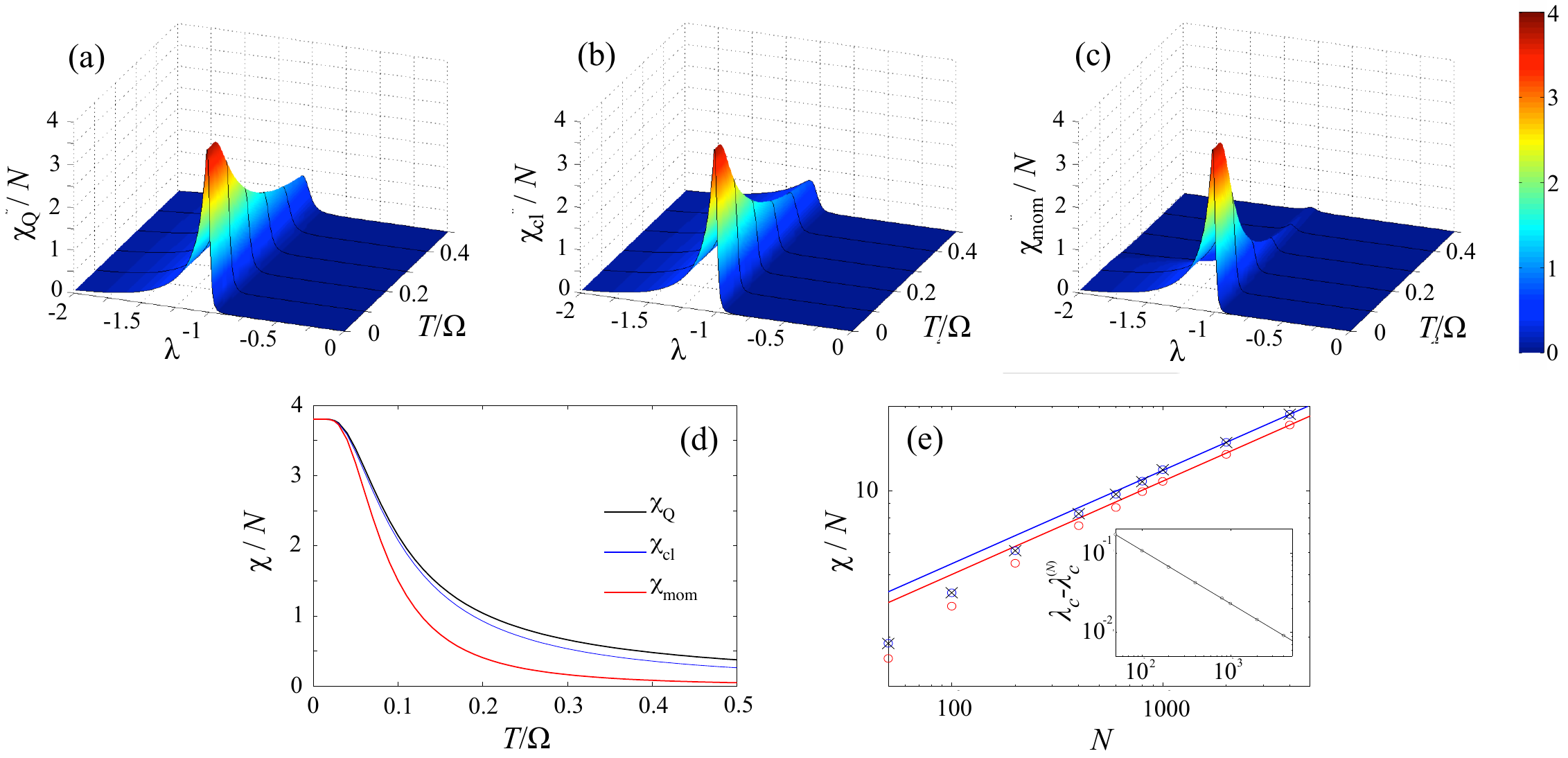}
\end{center}
\caption{Susceptibilities $\chi_{\rm Q}$ (a), $\chi_{\rm cl}$ (b) and $\chi_{\rm mom}$ (c) as a function of the temperature $T$ and the control parameter  $\lambda$. 
In panel (d) we plot $\chi_{\rm Q}$ (black line), $\chi_{\rm cl}$ (blue) and $\chi_{\rm mom}$ (red) 
as a function of $T/\Omega$ at $\lambda_c^{(N)}$.
In panels (a)-(d) $N=1000$ and $\delta/\Omega=2 \times 10^{-3}$.
In panel (e) we plot the optimized values of $\chi_{\rm Q}$ (black crosses), $\chi_{\rm cl}$ (blue circles) and $\chi_{\rm mom}$ (red circles)
as a function of $N$. 
Notice that $\chi_{\rm Q}$ agrees with $\chi_{\rm cl}$ in this case.
The solid lines are power-law fits, $\chi_{\rm Q}/N = \chi_{\rm cl}/N= 1.08 N^{1/3}$ (blue) and $\chimom/N = 1.18 N^{1/3}$ (red).
The inset shows $\lambda_c - \lambda_c^{(N)}$ as a function of $N$ (dots), 
the solid line is $\lambda_c - \lambda_c^{(N)} = 2.3 N^{-2/3}$. 
The numerical results confirm the expected scaling laws $\chimom/N \sim N^{1/3}$ and $\lambda_c - \lambda_c^{(N)} \sim N^{-2/3}$ (see text).}  
\label{Fig1}
\end{figure*} 

In the following, we study adiabatic sensing in a bosonic Josephson junction (BJJ).
We first consider the ideal system and then discuss its experimental realization.
The BJJ Hamiltonian is 
\be \label{BJJ}
\hat{H}_{\rm BJJ} = - \Omega \hat{J}_x + \zeta \hat{J}_z^2 + \delta \hat{J}_z,
\ee
where $\hat{J}_x = (\hat{a}^\dag \hat{b} + \hat{b}^\dag \hat{a})/2$, $\hat{J}_y = (\hat{a}^\dag \hat{b} + \hat{b}^\dag \hat{a})/(2i)$ and 
$\hat{J}_z = (\hat{a}^\dag \hat{a} - \hat{b}^\dag \hat{b})/2$ are angular momentum operators and
$\hat{a}$ and $\hat{b}$ are bosonic operators of the left and right reservoirs.
In Eq.~(\ref{BJJ}), $\Omega$ is the tunneling strength, $\zeta$ is the interaction strength and $\delta$ the energy imbalance. 
The BJJ model has a second-order symmetry-breaking QPT at a critical value $\lambda_c = -1$ of the control parameter 
$\lambda = N \zeta/\Omega$, in the thermodynamic limit 
$N\to \infty$, $\zeta/\Omega \to 0$ and for $\delta=0$.
The order parameter of the QPT is the population imbalance $\hat{J}_z$ between the $a$ and $b$ modes.
For $\lambda > \lambda_c$, the system is characterized by a paramagnetic quantum phase where $\mean{\hat{J}_z}=0$.
For $\lambda < \lambda_c$, we have a ferromagnetic phase where $\vert \mean{\hat{J}_z} \vert = (N/2)\sqrt{1-(\lambda_c/\lambda)}$ 
in the presence of symmetry breaking $\delta \neq 0$.
Numerical studies of the quantum fidelity susceptibility in the BJJ model~\cite{BuonsantePRA2012} and in 
the related Lipkin-Meshkov-Glick model~\cite{KwokPRE2008, SalvatoriPRA2014}
show that, at $T=0$, $\chiT(\lambda) = 1/[8(\lambda+1)^2]$ for $\lambda \gtrsim \lambda_c$. 
whereas $\chiT(\lambda) = N/[\vert \lambda \vert^3\sqrt{\lambda^2-1}]$ for $\lambda \lesssim \lambda_c$. 
At $\lambda=\lambda_c$ we have $\chiT(\lambda_c) \sim N^{4/3}$~\cite{KwokPRE2008, BuonsantePRA2012}, consistent with the prediction 
$\chiT(\lambda_c) \sim N^{2/(d\nu)}$, taking into account that $d=1$ and $\nu=3/2$ for this model.
Here we study $\chi_{\rm mom}$ and $\chicl$, Eqs.~(\ref{chimom}) and (\ref{chicl}), respectively, 
for the measurement of the order parameter $\hat{O} = \hat{J}_z$.
In Fig.~\ref{Fig1}(a)-(c) we compare $\chiT$, $\chicl$ and $\chimom$ as a function of $\lambda$ and the temperature $T$.
Specifically, we consider the equilibrium state 
$\rhoT(\lambda) = e^{-\beta \hat{H}(\lambda)} /\mathcal{Z(\lambda)}$, 
where $\mathcal{Z(\lambda)} = \tr[e^{-\beta \hat{H}(\lambda)}]$
is the partition function, $\beta=1/(k_BT)$, and $k_B$ is the Boltzmann constant.
The numerical study is based on the full diagonalization of the Hamiltonian (\ref{BJJ}) as a function of the control parameter $\lambda$, 
$\hat{H} \ket{\psi_n(\lambda)} = E_n(\lambda) \ket{\psi_n(\lambda)}$, where $\ket{\psi_n(\lambda)}$ and $E_n(\lambda)$ are eigenstates and 
eigenvectors, respectively.
Since we are interested in the behavior of the order parameter across the QPT, for the numerical analysis
we need to introduce a finite, small, symmetry-breaking term $\delta/\Omega \ll 1$. 
The three quantities $\chiT$, $\chicl$ and $\chimom$ show a clear peak close to $\lambda_c=-1$, 
thus giving a practical recipe to locate the critical point even at finite temperature.
In Fig.~\ref{Fig1}(d) we plot $\chiT$, $\chicl$ and $\chimom$ as a function of $T$ at $\lambda_c^{(N)}$.
For the parameters of the figure, the three quantities agree perfectly at $T=0$. 
At large temperature, $\chimom$ is dominated by $\chicl$ and $\chiT$, according to Eq.~(\ref{ineq}).
In Fig.~\ref{Fig1}(e) we study the scaling of $\chiT$, $\chicl$ and $\chimom$ as a function of $N$ at $\lambda_c^{(N)}$.
The results here are optimized as a function of $\delta$ such that $\chiT$, $\chicl$ and $\chimom$ have, as a function of $\lambda$, a maximum at $\lambda_c^{(N)}$
[regarding the values of $\lambda_c^{(N)}$, see the inset of Fig.~\ref{Fig1}(e), where we report the numerical $\lambda_c^{(N)}$ calculated 
as the value of $\lambda$ for which the energy gap $E_2 - E_0$ has a minimum].  
In particular, we recover the scaling behavior $\chimom(\lambda_c^{(N)}) \sim N^{4/3}$, predicted by Eq.~(\ref{scalingchimom1}).
$\chiT$ and $\chicl$ have the same scaling $N^{4/3}$, but with a slightly larger prefactor. 

\begin{figure}[h!]
\begin{center}
\includegraphics[clip,width=\columnwidth]{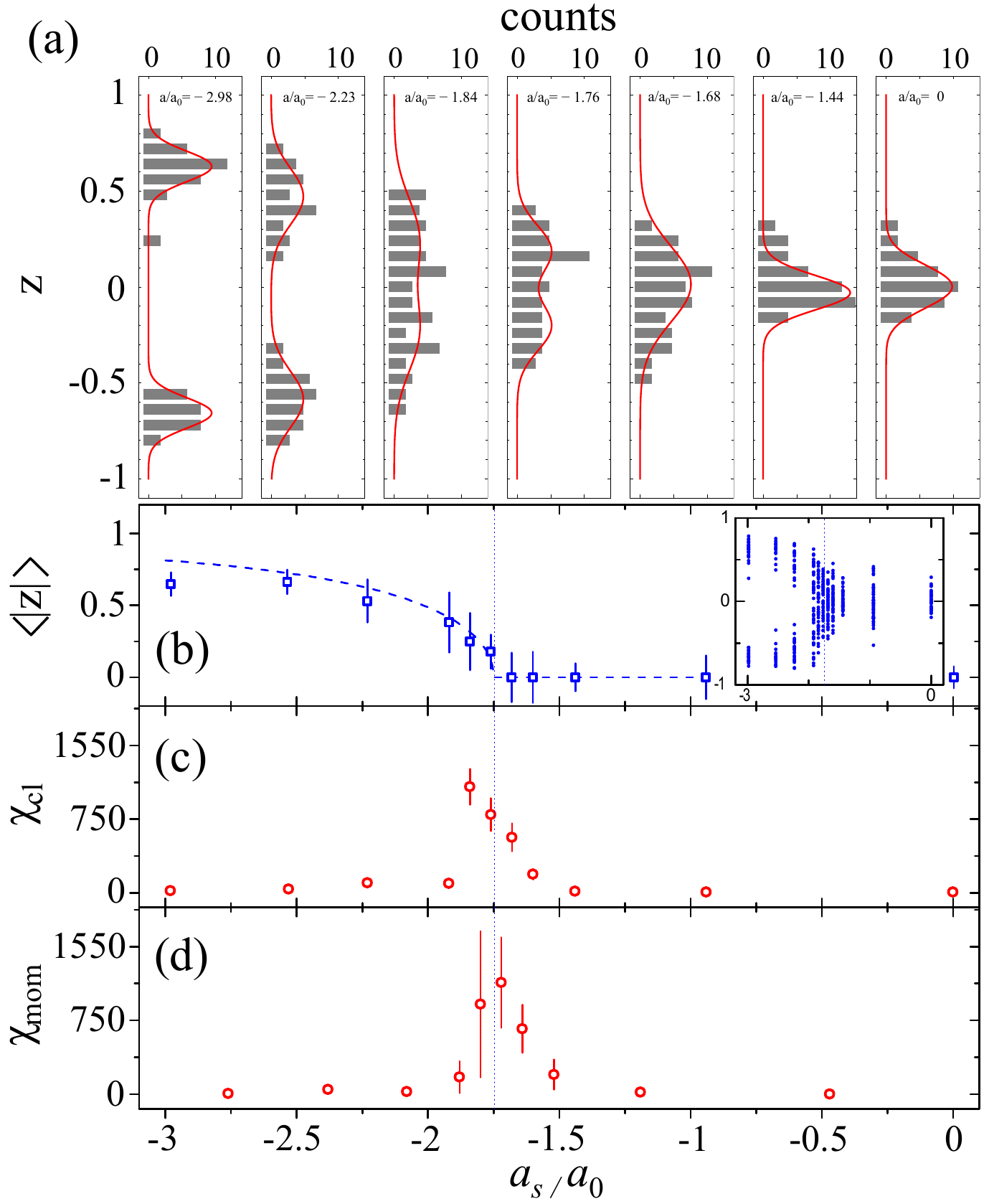}
\end{center}
\caption{(a) Histogram of measured $z$ values (gray, bin size 0.08) fitted with double Gaussian (red) for selected values of the scattering length. 
The inset of panel (b) shows the original $z$ values. 
(b) Mean value $\mean{|z|}$ as a function of the scattering length as obtained from histogram fit of panel (a). 
The error bars are the width of the individual Gaussian $\sigma_z$ obtained from the histogram fit. 
The dashed curve (blue) shows the model-dependent theoretical fitting function
$\mean{|z|} = \sqrt{1-(\lambda_c/\lambda)^2}$ used to extract the critical scattering length $a_c = -1.746(9)\,a_0$ -- highlighted as vertical dotted blue lines thru panels (b)-(d). 
Panels (c) and (d): $\chi_{\rm cl}$ and $\chi_{\rm mom}$ as a function of the scattering length. 
The error bars are the Gaussian $\sigma$ of a fit of the distribution of simulated $\chi_{\rm cl}$ or $\chi_{\rm mom}$ (see supplementary material). 
The position of the peaks of $\chi_{\rm cl}$ and $\chi_{\rm mom}$ agree very well with $a_c$.} 
\label{Fig2}
\end{figure}

Our experimental system consists of BJJ realized with a 
Bose-Einstein condensate of $^{39}$K atoms and confined in a double-well potential~\cite{TrenkwalderNATPHYS2016, SpagnolliPRL2017}.
The many-body Hamiltonian $\hat{H}  = \hat{H}_0 + \lambda \hat{H}_1$ is characterized by the competition between two energy terms:
$\hat{H}_0 = \int d \vect{r} \Psi^\dag (\vect{r}) [-\tfrac{\hbar^2}{2m} \nabla^2 + V(\vect{r})] \Psi (\vect{r})$ that includes kinetic and potential energy, and
$\hat{H}_1 = \tfrac{2\pi \hbar^2 a_0}{m}\int d \vect{r} \Psi^\dag (\vect{r}) \Psi^\dag (\vect{r}) \Psi (\vect{r}) \Psi (\vect{r})$ that
accounts for contact interaction between the atoms. 
Here, $\Psi (\vect{r})$ is the many-body wavefunction, $m$ is the atomic mass, $a_0$ is the Bohr radius, 
$V(\vect{r})$ the trapping potential (which is a double-well in the $x$ direction and harmonic in the orthogonal directions).
The control parameter is $\lambda = N a_s/a_0$, where $N$ is the number of atoms (here $N \approx 4500$) and $a_s$ the interatomic scattering length.
Crucially, in our experiment $a_s$ can be tuned to negative values corresponding to an attractive interaction between the atoms.
Notice that, within a two-mode approximation, $\Psi (\vect{r}) = \hat{a} \psi_a (\vect{r}) + \hat{b} \psi_b (\vect{r})$, where 
$\hat{a}$ and $\hat{b}$ are bosonic operators for the left and right well of the potential and $\psi_{a,b} (\vect{r})$
are mean-field wavefunction, the system Hamiltonian becomes Eq.~(\ref{BJJ})~\cite{MilburnPRA1997,SmerziPRL1997,JavanainenPRA1999}.
We emphasize that our analysis of the experimental results does not rely on any assumption, especially, they are not based on a two-mode approximation.
The system is prepared at a given value of $\lambda$ by adiabatically tuning the interaction strength while keeping the tunneling constant.
After state preparation, we directly measure the population $N_{R,L}$ of the right and left well, respectively, and report the 
population imbalance $z = (N_L - N_R)/(N_L+N_R)$.
At a critical value of the scattering length, the system undergoes a spontaneous symmetry-breaking QPT~\cite{TrenkwalderNATPHYS2016}.
For $\lambda < \lambda_c$ the population imbalance
becomes strongly sensitive to an uncontrolled energy mismatch (causing an asymmetry of the potential)
and one of the wells becomes more populated than the other. 
For different values of the scattering length we obtain a distribution histogram $P(z \vert \lambda)$, see Fig.~\ref{Fig2}(a). 
In Fig.~\ref{Fig2}(b) we plot $\mean{|z|}$ as a function of the scattering length. 
In a previous analysis of the experiment~\cite{TrenkwalderNATPHYS2016} 
a fit of $\langle |z| \rangle$ according to the theoretical fitting curve $\sqrt{1-(\lambda_c/\lambda)^2}$ for $\lambda < \lambda_c$ located the critical point $\lambda_c$. 
This fitting procedure is only justified at $T=0$.
Here, instead, we calculate the fidelity $\mathcal{F}_{\rm cl} (\lambda,\epsilon)$ between the histograms $P(z \vert \lambda)$, 
which is expected to converge relatively quickly to its infinite-sample limit~\cite{StrobelSCIENCE2014, PezzePNAS2016}.
Using $\epsilon$ from neighboring scattering lengths, a quadratic fit of 
$\mathcal{F}_{\rm cl}(\lambda,\epsilon) = 1 - \frac{1}{8} \chicl(\lambda) \epsilon^2 + O(\epsilon^3)$ gives access to $\chicl(\lambda)$.
The quantity $\chimom(\lambda)$ is accessed from the mean value and variance of $z$ as a function of $\lambda$:
\be
\chi_{\rm mom} = \frac{1}{(\sigma_z)^2} \bigg( \frac{d \mean{|z|}}{d \lambda} \bigg)^2,
\ee 
where $\mean{|z|}$ is given by half the distance between the peaks of a double Gaussian fitting the 
experimental histograms and $\sigma_z$ is the width of the individual Gaussian. The
derivative is obtained using the slope of $\mean{|z|}$ from neighboring scattering lengths, see supplementary material.
As expected, the experimental $\chicl$ [shown in Fig.~\ref{Fig2}(c)] and $\chimom$ [shown in Fig.~\ref{Fig2}(d)] 
have a similar behavior, as a function of scattering length:
they vary smoothly for $\lambda \neq \lambda_c$, while showing a 
clear peak at a critical value of $\lambda$. 
The peak allows to locate the critical point without any model-dependent fit to the data, 
or any assumption on temperature and technical noise.  
The peak of $\chicl$ and $\chimom$ corresponds to the point where the symmetry breaking occurs, 
in agreement with the results of Ref.~\cite{TrenkwalderNATPHYS2016}.
According to the close relation between sensitivity and susceptibility, a maximum of $\chi_{\rm mom}$, or $\chi_{\rm cl}$, 
witnesses as a point of minimum uncertainty $\Delta \lambda$ and maximum sensitivity in the estimation of the parameter $\lambda$
and thus certifies that quantum criticality can enhance the sensitivity of measurements.
The  results show that $\chi_{\rm mom} \approx \chi_{\rm cl}$ are about an order of magnitude smaller than the 
theoretical predictions of the two-mode model at zero temperature (we estimate a temperature $T \approx 10$ nK. 
This suggests that the experimental limiting factor is given by the residual uncontrolled fluctuations of the energy imbalance  
between the two wells.

In conclusion, we have experimentally demonstrated that 
the estimation sensitivity of the control parameter in an adiabatic sensor is enhanced around the critical point of a QPT.
Furthermore, our methods have allowed to locate, at finite temperature, the critical point of a QPT, 
based on the measurement of the relative number of particles 
between the two wells and without any model-dependent theoretical fit to the data. 
Reducing the temperature and fluctuations of the double well potential may give access to a regime where 
$\chicl$ and $\chimom$ have a super-extensive scaling with $N$ and, equivalently, 
$(\Delta \lambda)^2$ scales faster than $1/N$.
Our methods and results are a guideline for sensing applications and 
precise location of QPTs in complex many-body systems such as trapped-ions quantum simulators~\cite{BrittonNATURE2012, ZhangNATURE2017, BernienNATURE2017} 
ultracold atoms platforms~\cite{BaumannNATURE2010, SimonNATURE2011,EndersNATURE2012, IslamNATURE2015},
and electronic materials~\cite{ColdeaSCIENCE2010,KinrossPRX2014}.  \\
   
\begin{acknowledgments}   

{\it Acknowledgments.} We thank Augusto Smerzi for discussion.

\end{acknowledgments}


\newpage

\section{Supplementary Information}

For the analysis, we have used the data of Ref.~\cite{TrenkwalderNATPHYS2016}, see the inset of Fig.~\ref{Fig2}b, obtained with 
tunneling rate $\Omega = 40$ Hz, atom number $N \approx 4500$, temperature $\approx$ 10\,nK.

\subsection{Data analysis of \Chimom{}}

According to Eq.~(\ref{chimom}),
\begin{equation}\label{eq:chimom-def}
\chi_{\rm mom}(\lambda) = \frac{1}{(\Delta \hat{O})^2} \bigg( \frac{d \big\langle \hat{O} \big\rangle}{d \lambda} \bigg)^2  
\equiv a_0^2 \left(\frac{1}{\sigma}\frac{d |z|}{d a_s}\right)^2 \ ,
\end{equation}
with $\lambda \equiv a_s/a_0$ with $a_s$ the (s-wave) scattering length and $a_0$ the Bohr radius and the observable $\langle \hat O \rangle \equiv |z|$ with $|z| = \frac{|N_L-N_R|}{N_R+N_L}$ the splitting of the number of atoms in the left and right well, $N_L$ and $N_R$ respectively. 
The variance of the observable $(\Delta \hat O )^2 \equiv \sigma$ is obtained from the fit of two Gaussian on the measured distribution of observed $z$ values 
(for details see Ref.~\cite{TrenkwalderNATPHYS2016}). 
The splitting $|z|$ is half of the distance between the two Gaussians and $\sigma$ is the width of each of the Gaussians. 
Sample distributions and fitted Gaussians are plotted in Fig.~\ref{Fig2}a and the resulting $|z|$ is shown in Fig.~\ref{Fig2}b. 
Applying Eq.~\eqref{eq:chimom-def} directly to the data gives large errors since we must obtain the derivative of discrete and noisy data. 
To avoid this and to improve the quality of the result without the need of excessive smoothing of the (limited) 
data we simulate the experiment and analysis using the double Gaussian fit of the observed distribution as the sample distribution. 
On each of the about 3000 simulations we calculate \Chimom{} with Eq.~\eqref{eq:chimom-def} using the derivative of nearest neighbors 
in $a_s$ and collect the result in histograms for each scattering length $a_s$, see Fig.~\ref{FigSupp1}. 
We find that the histograms can be best fitted with a Gaussian (red) on an exponential background centered at \Chimom{} = 0. 
For the data point at $a_s = -2.76\,a_0$ the fit with the background gives \Chimom $\approx 0$ but with a very big error, when we fit without 
background we get still a result close to zero but with a much smaller error. 
We fit the histograms all in linear scale since this emphasizes the large count rates of \Chimom. 
However, fitting in logarithmic scale does not change much the result, except for the data point at $a_s = -1.88\, a_0$, see second row of Fig.~\ref{FigSupp1}. 
This data point is difficult to assign but we believe that the fit in linear scale (left in the figure) is better describing the histogram for large count rates than the fit done in logarithmic scale (right in the figure). The results shown in Fig.~\ref{Fig2}d report the center of the fitted Gaussian as the resulting \Chimom{} and the width of the Gaussian as the error. \\

\subsection{Data analysis of \Chicl{}}

The classical fidelity susceptibility \Chicl{} is defined in Eq.~(\ref{chicl}) from the fidelity $\mathcal{F}_{\rm cl} (\lambda,\epsilon) = \sum_{\mu} \sqrt{P(\mu \vert \lambda) P(\mu \vert \lambda+\epsilon) }$ between neighbouring probability distributions $P(\mu \vert \lambda)$ and $P(\mu \vert \lambda+\epsilon)$ and $\lambda \equiv \frac{a_s}{a_0}$ and $\epsilon \rightarrow 0$. Expanding $\mathcal{F}_{\rm cl} (\lambda,\epsilon)$ to second order in $\epsilon$ gives: 
\be
\mathcal{F}_{\rm cl}(\lambda,\epsilon) = 1 - \frac{\chi_{cl}}{8} \epsilon^2 + O(\epsilon^3).
\ee
To estimate \Chicl{} from our data we take for each scattering length $a^i_s$ the distribution $P(\mu \vert \lambda_i) \equiv h_i(\mu)$ from the original data $\mu \equiv z$ with a fixed bin size $\delta z = 0.05$. Each distribution is normalized such that $\sum_z h_i(z) = 1$. We approximate the overlap $\mathcal{F}_{\rm cl}(\lambda,\epsilon)$ by the overlap of two neighboring distributions:
\begin{equation}
\begin{array}{ll}
& \mathcal{F}_{\rm cl}(\lambda_i,\epsilon_{ij}) \approx \sum_z \sqrt{h_i(z) h_j(z)} \\
& \epsilon_{ij} = \lambda_{j}-\lambda_{i} \equiv \frac{a_j - a_i}{a_0}\ .
\end{array}
\end{equation}
For each $\lambda_i$ we take the nearest neighboring points $\{x_j, y_j\} = \{\epsilon_{ij},1-\mathcal{F}_{\rm cl,ij}\}$ with $j = i \pm 1$ 
and fit a parabola through them, where the only free parameter $\chi_{cl,i}$ is obtained from the fit. Note, that for \Chimom{} 
we have fitted the original distribution with a double Gaussian, which for the direct analysis of \Chicl{} would not be needed. 
However, to improve the quality of the result and to obtain errors we again proceed as for the analysis of \Chimom{}: 
we generate 3000 realizations of the experiment using the double Gauss fit of the original data as the distribution 
of the samples, and we calculate \Chicl{} for each sample and collect the result in histograms,
see Fig.~\ref{FigSupp2}. Each resulting histogram is very well fit by a single Gaussian (red) without background. 
The results shown in Fig.~\ref{Fig2}c report the center of the Gaussian as the resulting \Chicl{} and the Gaussian $\sigma$ as the error.

\begin{widetext}

\begin{figure}[ht]
\includegraphics[width=\columnwidth] {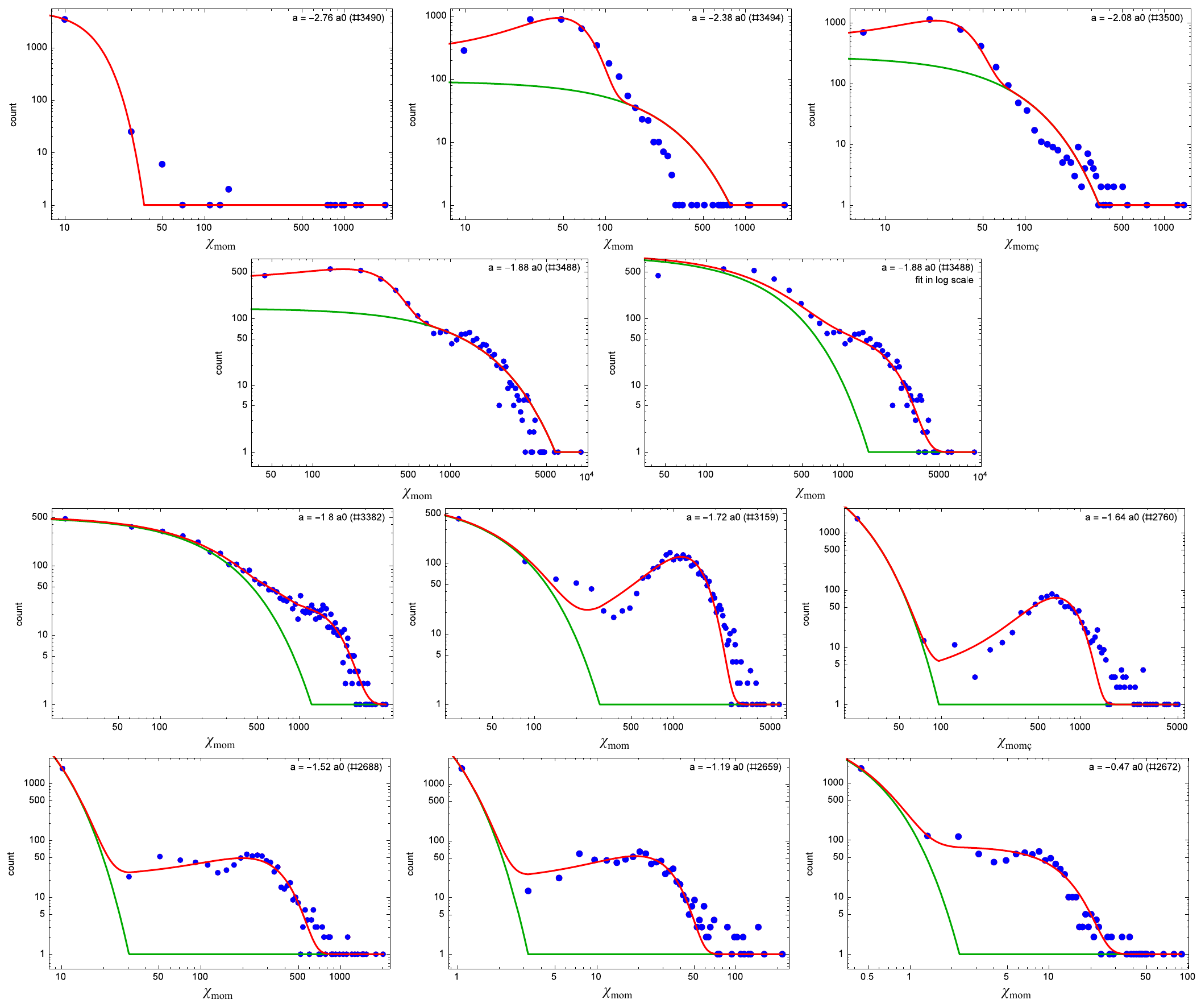}
\caption{\label{fig:chi-mom-var}Histograms of \Chimom{} (blue dots, 100 bins) generated from about 3000 simulations (number after ``\#'' in label) of random samples obtained from the distribution described by the double Gaussian fit of the original data. Each histogram is fitted with a Gaussian (red) on an exponential background (green, limited to 1), except for $a_s = -2.76\,a_0$ fitted without background. The data for $a_s = -1.88\,a_0$ is shown twice (second row) where we show the fit result when the histogram is fitted in linear scale (left), as was done for all other histograms, and for comparison the fit result when the histogram is fitted in logarithmic scale (right).}
\label{FigSupp1}
\end{figure}

\begin{figure}[ht]
\includegraphics[width=\columnwidth] {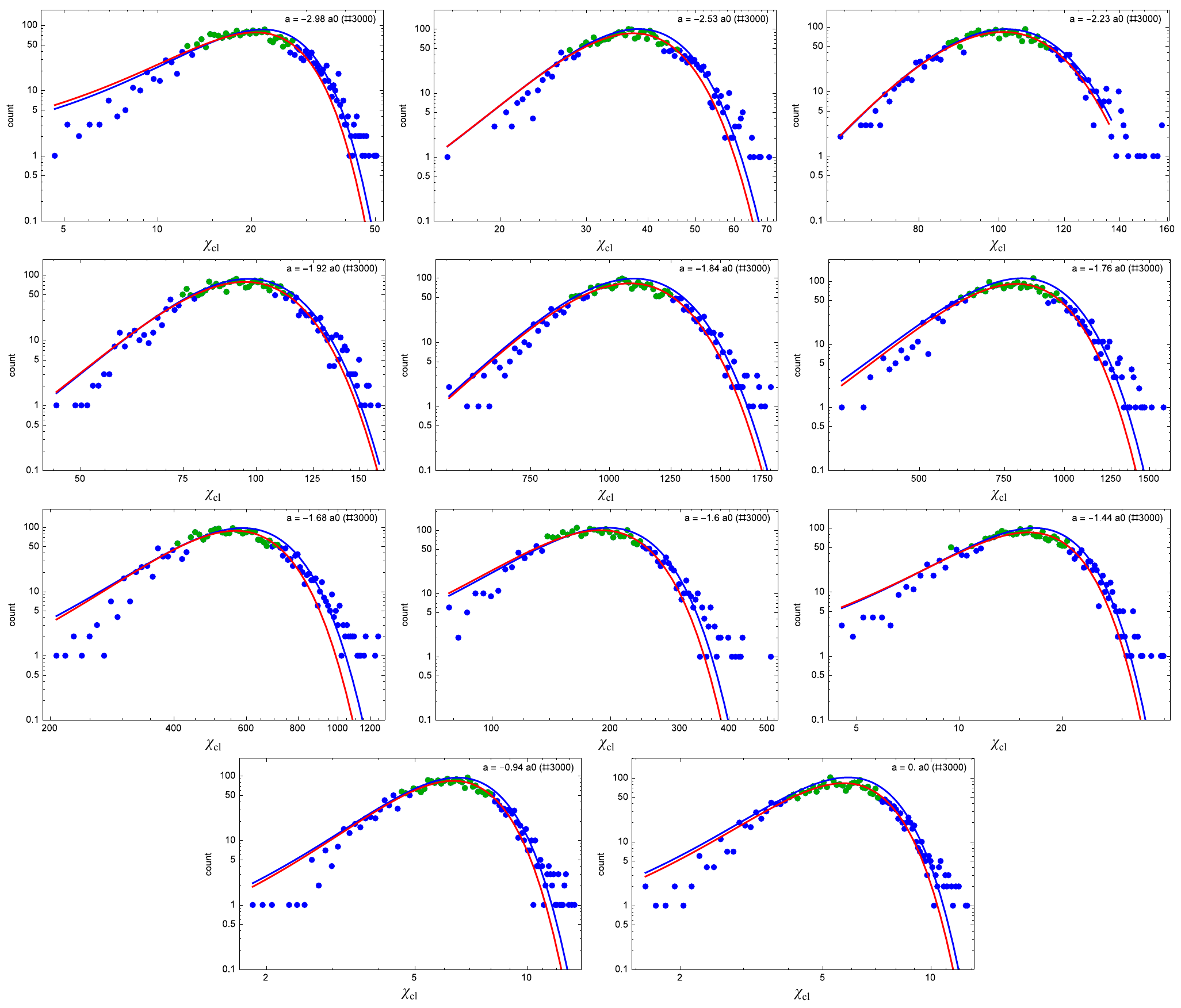}
\caption{\label{fig:chi-cl-var}Histograms of \Chicl{} (blue and green dots, 100 bins) generated from 3000 simulations (number after ``\#'' in label) of random samples obtained from the distribution described by the double Gaussian fit of the original data. Each histogram is fitted with a simple Gaussian (red). For comparison we also show the mean value and standard deviation (blue Gaussian) and the 68\,\% confidence region (green dots).}
\label{FigSupp2}
\end{figure}

\end{widetext}


\begin{thebibliography}{400}

\bibitem{PezzeRMP2018}
L. Pezz\`e, A. Smerzi, M. K. Oberthaler, R. Schmied and P. Treutlein, 
``Quantum metrology with nonclassical states of atomic ensembles,''
Rev. Mod. Phys. {\bf 90}, 035005 (2018).

\bibitem{DeganRMP2017}
C. L. Degen, F. Reinhard, and P. Cappellaro, 
``Quantum sensing,'' 
Rev. Mod. Phys. {\bf 89}, 035002 (2017).

\bibitem{SchnabelNATCOM2010}
R. Schnabel, N. Mavalvala, D. E. McClelland, and P. K. Lam, 
``Quantum metrology for gravitational wave astronomy'', 
Nat. Comm. {\bf 1}, 121 (2010).

\bibitem{LudlowRMP2015}
A. S. Ludlow, M.M. Boyd, J. Ye, E. Peik, and P. O. Schmidt,
``Optical atomic clocks'',
Rev. Mod. Phys. {\bf 87}, 637 (2015).

\bibitem{TaylorPR2016}
M. A. Taylor and W. P. Bowen, 
``Quantum metrology and its application in biology'', 
Phys. Rep. {\bf 615}, 1 (2016)

\bibitem{ZanardiPRA2008}
P. Zanardi, M.G.A. Paris, and L. Campos Venuti, 
``Quantum criticality as a resource for quantum estimation'', 
Phys. Rev. A {\bf 78}, 042105 (2008).

\bibitem{SachdevBOOK}
S. Sachdev, 
{\it Quantum Phase Transitions} 
(Cambridge University Press, Cambridge, 1999).

\bibitem{DuttaBOOK}
A. Dutta, G. Aeppli, B. K. Chakrabarti, U. Divakaran, T. F. Rosenbaum and D. Sen, 
{\it Quantum Phase Transitions in Transverse Field Spin Models: From Statistical Physics to Quantum Information} 
(Cambridge University Press, Cambridge, 2015)

\bibitem{ZanardiPRE2006}
P. Zanardi and N. Paunkovi\'c, 
``Ground State Overlap and Quantum Phase Transitions'', 
Phys. Rev. E {\bf 74}, 031123 (2006).

\bibitem{YouPRE2007}
W.-L. You, Y.-W. Li, and S.-J. Gu, 
``Fidelity, dynamic structure factor, and susceptibility in critical phenomena'', 
Phys. Rev. E {\bf 76}, 022101 (2007).

\bibitem{ZanardiPRL2007}
P. Zanardi, P. Giorda and M. Cozzini, 
``Information-Theoretic Differential Geometry of Quantum Phase Transitions'',
Phys. Rev. Lett. {\bf 99}, 100603 (2007).

\bibitem{GuIJMPB2010} 
S.-J. Gu, 
``Fidelity Approach to Quantum Phase Transitions'', 
Int. J. Mod. Phys. B {\bf 24}, 4371 (2010).

\bibitem{BraunRMP2018}
D. Braun, G. Adesso, F. Benatti, Ro. Floreanini, U. Marzolino, M. W. Mitchell, and S. Pirandola,
``Quantum-enhanced measurements without entanglement'', 
Rev. Mod. Phys. {\bf 90}, 035006 (2018).

\bibitem{CamposVenutiPRL2007}
L. Campos Venuti and P. Zanardi,
``Quantum Critical scaling of the Geometric Tensor'',
Phys. Rev. Lett. {\bf 99}, 096701 (2007).

\bibitem{SchwandtPRL2009}
D. Schwandt, F. Alet, and S. Capponi, 
``Quantum critical scaling of fidelity susceptibility'', 
Phys. Rev. Lett. {\bf 103}, 170501 (2009);
A.F. Albuquerque, F. Alet, C. Sire, and S. Capponi, 
``Quantum critical scaling of fidelity susceptibility''
Phys. Rev. B {\bf  81}, 064418 (2010).

\bibitem{PolkovnikovBOOK}
A. Polkovnikov and V. Gritsev, 
``Universal Dynamics Near Quantum Critical Points'', in {\it Understanding Quantum Phase Transitions}, 
edited by Lincoln D. Carr (Taylor \& Francis, Boca Raton, 2010).

\bibitem{CozziniPRB2007}
M. Cozzini, P. Giorda, and Paolo Zanardi,
``Quantum phase transitions and quantum fidelity in free fermion graphs'',
Phys. Rev. B {\bf 75}, 014439 (2007).

\bibitem{BuonsantePRL2007}
P. Buonsante and A. Vezzani
``Ground-State Fidelity and Bipartite Entanglement in the Bose-Hubbard Model'', 
Phys. Rev. Lett. {\bf 98}, 110601 (2007).

\bibitem{InvernizziPRA2008}
C. Invernizzi, M. Korbman, L. Campos Venuti, and M.G.A. Paris, 
``Optimal quantum estimation in spin system at criticality'', 
Phys. Rev. A {\bf 78}, 042106 (2008).

\bibitem{KwokPRE2008}
H.M. Kwok, W.Q. Ning, S.J. Gu, and H.Q. Lin, 
``Quantum criticality of the Lipkin-Meshkov-Glick model in terms of fidelity susceptibility'',
Phys. Rev. E {\bf 78}, 032103 (2008).

\bibitem{RamsPRL2011}
M.M. Rams and B. Damski, 
``Quantum Fidelity in the Thermodynamic Limit'',
Phys. Rev. Lett. {\bf 106}, 055701 (2011).

\bibitem{BuonsantePRA2012}
P. Buonsante, R. Burioni, E. Vescovi, and A. Vezzani,
``Quantum criticality in a bosonic Josephson junction'',
Phys. Rev. A {\bf 85}, 043625 (2012). 

\bibitem{SalvatoriPRA2014}
G. Salvatori, A. Mandarino, and M.G.A. Paris, 
``Quantum metrology in Lipkin-Meshkov-Glick critical systems'', 
Phys. Rev. A {\bf 90}, 022111 (2014).

\bibitem{WangPRX2015}
L. Wang, Y.-H. Liu, J. Imriska, P. N. Ma, and M. Troyer,
``Fidelity Susceptibility Made Simple: A Unified Quantum Monte Carlo Approach'',
Phys. Rev. X {\bf 5}, 031007 (2015).

\bibitem{MehboudiPRA2016}
M. Mehboudi, L.A. Correa, and A. Sanpera, 
``Achieving sub-shot-noise sensing at finite temperature'', 
Phys. Rev. A {\bf 94}, 042121 (2016).

\bibitem{RamsPRX2018}
M.M. Rams, P. Sierant, O. Dutta, O. Horodecki, and J. Zakrewski, 
``At the Limits of Criticality-Based Quantum Metrology: Apparent Super-Heisenberg Scaling Revisited'', 
Phys. Rev. X {\bf 8}, 021022 (2018).

\bibitem{GuEPL2014}
S.-J. Gu and W. C. Yu,
``Spectral function and fidelity susceptibility in quantum critical phenomena'',
Europhys. Lett. {\bf 108} 20002 (2014).

\bibitem{ZhangPRL2008}   
J. Zhang, X. Peng, N. Rajendran, and D. Suter,
``Detection of Quantum Critical Points by a Probe Qubit'',
Phys. Rev. Lett. {\bf 100}, 100501 (2008).

\bibitem{ZhangPRA2009}
J. Zhang, F.M. Cucchietti, C.M. Chandrashekar, M. Laforest, C.A. Ryan, M. Ditty, A. Hubbard, J.K. Gamble, and R. Laflamme,
``Direct observation of quantum criticality in Ising spin chains'',
Phys. Rev. A {\bf 79}, 012305 (2009).

\bibitem{TrenkwalderNATPHYS2016}
A. Trenkwalder, G. Spagnolli, G. Semeghini, S. Coop, M. Landini, P. Castilho, L. Pezz\`e, G. Modugno, M. Inguscio, A. Smerzi, and M. Fattori, 
``Quantum phase transitions with parity-symmetry breaking and hysteresis'',
Nat. Phys. {\bf 12}, 826 (2016).

\bibitem{SpagnolliPRL2017}
G. Spagnolli, G. Semeghini, L. Masi, G. Ferioli, A. Trenkwalder, S. Coop, M. Landini, L. Pezz\`e, G. Modugno, M. Inguscio, A. Smerzi, and M. Fattori
``Crossing Over from Attractive to Repulsive Interactions in a Tunneling Bosonic Josephson Junction'', 
Phys. Rev. Lett. {\bf 118}, 230403 (2017).

\bibitem{footnote1}
Besides adiabatic sensing discussed in this manuscript, 
the generation of metrological entanglement created at critical points of QPTs is 
under intense investigation~\cite{MaPRA2009, HaukeNATPHYS2016, PezzePRL2017, FrerotPRL2018, GabbrielliSCIREP2018, GabbrielliNJP2019}.

\bibitem{MaPRA2009}
J. Ma, and X. Wang, 
``Fisher information and spin squeezing in the Lipkin-Meshkov-Glick model'', 
Phys. Rev. A {\bf 80}, 012318 (2009).

\bibitem{HaukeNATPHYS2016}
P. Hauke, M. Heyl, L. Tagliacozzo, and P. Zoller, 
``Measuring multipartite entanglement through dynamic susceptibilities'', 
Nat. Phys. {\bf 12}, 778 (2016).

\bibitem{PezzePRL2017}
L. Pezz\`e, M. Gabbrielli, L. Lepori, and A. Smerzi, 
``Multipartite Entanglement in Topological Quantum Phases'', 
Phys. Rev. Lett. {\bf 119}, 250401 (2017);
Y.-R. Zhang, Y. Zeng, H. Fan, J. Q. You, and F. Nori,
``Characterization of Topological States via Dual Multipartite Entanglement'', 
Phys. Rev. Lett. {\bf 120}, 250501 (2018).

\bibitem{FrerotPRL2018}
I. Fr\'erot and T. Roscilde, 
``Quantum Critical Metrology'', 
Phys. Rev. Lett. {\bf 121}, 020402 (2018).

\bibitem{GabbrielliSCIREP2018}
M. Gabbrielli, A. Smerzi, and L. Pezz\`e,
``Multipartite entanglement at finite temperature'', 
Sci. Rep. {\bf 8}, 15663 (2018);
I. Fr\'erot and T. Roscilde,
``Reconstructing the quantum critical fan of strongly correlated systems using quantum correlations'', 
Nat. Comm. {\bf 10}, 577 (2019).

\bibitem{GabbrielliNJP2019}
M. Gabbrielli, L. Lepori, and L. Pezz\`e
``Multipartite entanglement tomography of a quantum simulator''
New J. Phys. {\bf 21} 033039 (2019).

\bibitem{PezzeBOOK}
L. Pezz\`e and A. Smerzi, 
``Quantum theory of phase estimation'', in Atom Interferometry, Proceedings of the International School of Physics ``Enrico Fermi'', 
Course 188, Varenna, edited by G. M. Tino and M. A. Kasevich (IOS Press, Amsterdam) p. 691 (2016).

\bibitem{CramerBOOK}
H. Cram\`er, 
{\it Mathematical Methods of Statistics} (Princeton University Press, Princeton, NJ, 1946).

\bibitem{HelstromBOOK}
C.W. Helstrom, 
{\it Quantum Detection and Estimation Theory} (Academic Press, New York, 1976).

\bibitem{HelstromPLA1967}
C.W. Helstrom, 
``Minimum mean-squared error of estimates in quantum statistics'', 
Phys. Lett. A {\bf 25}, 101 (1967).

\bibitem{BraunsteinPRL1994}
S.L. Braunstein and C.M. Caves, 
``Statistical Distance and the Geometry of Quantum States'', 
Phys. Rev. Lett. {\bf 72}, 3439 (1994).

\bibitem{UhlmannRMP1976}
A. Uhlmann, 
``The ``transition probability'' in the state space of a *-algebra'',
Rep. Math. Phys. {\bf 9}, 273 (1976)

\bibitem{JozsaJMO1994}
R. Jozsa, 
``Fidelity for mixed quantum states'', 
J. Mod. Opt. {\bf 41}, 2315 (1994).

\bibitem{nota5}
There is always an optimal measurement -- given by the projection over the eigenstates of the operator $\hat{L}(\lambda)$ 
defined as $2 \partial_\lambda \hat{\rho}(\lambda) = \hat{L}(\lambda) \hat{\rho}(\lambda) + \hat{\rho}(\lambda)  \hat{L}(\lambda)$ -- 
such that the equality $\chiT(\lambda) = \chicl(\lambda)$ is saturated (at zero as well as at finite temperature)~\cite{BraunsteinPRL1994}.
Yet, this observable, that depends on the equilibrium state of the system $\hat{\rho}(\lambda)$ and 
its derivative $\partial_\lambda \hat{\rho}(\lambda)$ is difficult to implement in practice.

\bibitem{ContentinoBOOK}
M.A. Contentino,
{\it Quantum scaling in many-body systems} 
(World Scientific Publishing, Singapore, 2001).

\bibitem{PezzePNAS2016}    
L. Pezz\`e, Y. Li, W. Li, and A. Smerzi, 
``Witnessing entanglement without entanglement witness operators,''
Proceedings of the National Academy of Sciences {\bf 113}, 11459 (2016).    

\bibitem{StrobelSCIENCE2014}
H. Strobel, W. Muessel, D. Linnemann, T. Zibold, D. B. Hume, L. Pezz\`e, A. Smerzi, and M. K. Oberthaler,
``Fisher information and entanglement of non-Gaussian spin states,'' 
Science {\bf 345}, 424 (2014).

\bibitem{MilburnPRA1997}
G.J. Milburn, J. Corney, E.M. Wright, and D.F. Walls, 
``Quantum dynamics of an atomic Bose-Einstein condensate in a double-well potential,''
Phys. Rev. A {\bf 55}, 4318 (1997).

\bibitem{SmerziPRL1997}
A. Smerzi, S. Fantoni, S. Giovanazzi, and S.R. Shenoy, 
``Quantum coherent atomic tunneling between two trapped Bose- Einstein condensates,''
Phys. Rev. Lett. {\bf 79}, 4950 (1997).

\bibitem{JavanainenPRA1999}
J. Javanainen and M.Yu. Ivanov, 
``Splitting a trap containing a Bose-Einstein condensate: Atom number fluctuations,'' 
Phys. Rev. A {\bf 60}, 2351 (1999).

\bibitem{BrittonNATURE2012}
J.W. Britton, B.C. Sawyer, A.C. Keith, C.J. Wang, J.K. Freericks, H. Uys, M.J. Biercuk and J.J. Bollinger, 
``Engineered two-dimensional Ising interactions in a trapped-ion quantum simulator with hundreds of spins'', 
Nature {\bf 484}, 489 (2012).

\bibitem{ZhangNATURE2017}
J. Zhang, G. Pagano, P. W. Hess, A. Kyprianidis, P. Becker, H. Kaplan, A.V. Gorshkov, Z.-X. Gong, and C. Monroe, 
``Observation of a many-body dynamical phase transition with a 53-qubit quantum simulator'', 
Nature {\bf 551}, 601 (2017).

\bibitem{BernienNATURE2017}
H. Bernien, S. Schwartz, A. Keesling, H. Levine, A. Omran, H. Pichler, S. Choi, A. S. Zibrov, M. Endres, M. Greiner, V. Vuletic, and M. D. Lukin, 
``Probing many-body dynamics on a 51-atom quantum simulator'', 
Nature {\bf 551}, 579 (2017).

\bibitem{BaumannNATURE2010}
K. Baumann, C. Guerlin, F. Brennecke, and T. Esslinger, 
``Dicke quantum phase transition with a superfluid gas in an optical cavity'', 
Nature {\bf 464}, 1301 (2010).

\bibitem{SimonNATURE2011}
J. Simon, W.S. Bakr, R. Ma, M.E. Tai, P.M. Preiss and M. Greiner, 
``Quantum simulation of antiferromagnetic spin chains in an optical lattice'', 
Nature {\bf 472}, 307 (2011).

\bibitem{EndersNATURE2012}
M. Endres, T. Fukuhara, D. Pekker, M. Cheneau, P. Schauss, C. Gross, E. Demler, S. Kuhr, and I. Bloch,
``The Higgs amplitude mode at the two-dimensional superfluid/Mott insulator transition'', 
Nature {\bf 487}, 545 (2012).

\bibitem{IslamNATURE2015}
R. Islam, R. Ma, P. M. Preiss, M. E. Tai, A. Lukin, M. Rispoli, and Markus Greiner,
``Measuring entanglement entropy in a quantum many-body system'',
Nature {\bf 528}, 77 (2015)

\bibitem{ColdeaSCIENCE2010}
R. Coldea, D. A. Tennant, E. M. Wheeler, E. Wawrzynska, D. Prabhakaran, M. Telling, K. Habicht, P. Smeibidl, and K. Kiefer, 
``Quantum Criticality in an Ising Chain: Experimental Evidence for Emergent $E_8$ Symmetry'', 
Science {\bf 327}, 177 (2010).

\bibitem{KinrossPRX2014}
A. W. Kinross, M. Fu, T. J. Munsie, H. A. Dabkowska, G. M. Luke, S. Sachdev, and T. Imai,
``Evolution of Quantum Fluctuations Near the Quantum Critical Point
of the Transverse Field Ising Chain System CoNb$_2$O$_6$'', 
Phys. Rev. X {\bf 4}, 031008 (2014).

\end{thebibliography}
\end{document}